\documentclass{article}

\usepackage{multirow}

\usepackage{arxiv}

\usepackage[utf8]{inputenc} 
\usepackage[T1]{fontenc}    
\usepackage{hyperref}       
\usepackage{url}            
\usepackage{booktabs}       
\usepackage{amsfonts}       
\usepackage{nicefrac}       
\usepackage{microtype}      
\usepackage{lipsum}		
\usepackage{graphicx}
\graphicspath{{media/}} 
\usepackage[numbers]{natbib}
\usepackage{doi}

\title{Assessment of the Financial Competitiveness of Publicly Listed Indian Real Estate Companies Using the Entropy Method}

\author{ \href{https://orcid.org/0009-0001-7169-1617}{\includegraphics[scale=0.06]{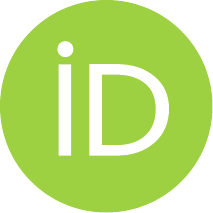}\hspace{1mm}Ritij Saini} \\
	Aurum PropTech Limited \\
	\texttt{ritij.saini@aurumproptech.in} \\
	\And
	Aditya Deora \\
	Aurum PropTech Limited \\
	\texttt{aditya.deora@aurumproptech.in} \\
	\AND
	 \\
	 \\
	 \\
	\texttt{} \\
	\And
	Kirtesh Gadiya \thanks{Prime Minister Research Fellow, Department of Civil Engineering, IIT Bombay} \\
	IIT Bombay \\
	\texttt{kirtesh\_gadiya@iitb.ac.in} \\
	\And
	 \\
	 \\
	 \\
}

\hypersetup{
pdftitle={Assessment of the Financial Competitiveness of Publicly Listed Indian Real Estate Companies Using the Entropy Method},
pdfsubject={q-fin.GN},
pdfauthor={Ritij Saini, Aditya Deora},
pdfkeywords={Real Estate, PropTech, Financial Analysis, Indian Listed Companies, Evaluation, Entropy},
}

\begin{document}
\maketitle

\begin{abstract}
	The real estate sector is one of the key drivers of India's national economy, contributing about 7.3\%  to the GDP. As the market evolves, more players enter, and government policies become more stringent, Indian real estate companies face increasing competition. Improving financial competitiveness is crucial for the survival and growth of these companies. This paper presents a financial competitiveness evaluation index system for Indian-listed real estate companies, covering profitability, solvency, and operational capacity. Using key financial ratios and a scoring system, the financial competitiveness of various companies was evaluated, revealing that companies with high scores have strong profitability and operational capacity. In contrast, those with lower scores struggle with solvency and working capital. 
\end{abstract}

\keywords{Real Estate \and PropTech \and Financial Analysis \and Indian Listed Companies \and Evaluation \and Entropy}

\section{Introduction}

Over the past two decades, the Indian real estate industry has experienced significant growth and has become a major pillar of the economic growth of the Indian subcontinent. However, the sharp rise in property prices has resulted in slow demand nationwide \cite{tamila}. In response, the Indian government has introduced several regulatory measures to control property prices. Notably, in 2016, the implementation of the Real Estate (Regulation and Development) Act (RERA) \cite{rera} aimed to bring more transparency and accountability to the sector. Additionally, introducing the Goods and Services Tax (GST) \cite{gst} in 2017 and demonetization in 2016 have impacted the real estate market.

These regulations have begun to show their effects. Restrictions on speculative investments and transaction-related challenges have pressured real estate companies, slowing down inventory turnover and tightening cash flows. Companies are now dealing with increased capital costs and heightened financial risk. To survive and grow, real estate developers must enhance their competitiveness since real estate investment in India is capital-intensive, and many developers carry high debt loads. As a result, financial competitiveness has become a crucial factor for the survival and growth of these companies. Securing financing is vital in determining a company's long-term success in the real estate market.

India's real estate sector has grown significantly over the past decade, supported by increasing urbanization, rising middle-class incomes, and government incentives. However, with changes in market dynamics, real estate companies in India now face heightened competition and stricter regulations. To stay competitive, companies must continuously improve their financial performance. This paper evaluates the financial competitiveness of Indian-listed real estate companies using profitability, solvency, and operational capacity metrics. The findings will provide insights into the financial health of these companies, helping them navigate the challenges posed by market volatility and regulatory shifts.

\section{Literature Review}
\label{sec:headings}

Since Stephen H. Hymer introduced the concept of enterprise competitiveness in his 1960 PhD dissertation, "The International Operations of National Firms: A Study of Direct Foreign Investment", the topic has attracted significant attention from academia and industry. In the context of real estate companies, financial competitiveness has garnered significant attention, especially in highly regulated and capital-intensive markets such as China and India. Chinese real estate firms \cite{lin2013evaluationfinancialcompetitivenesschinese}, underscores the importance of profitability, solvency, and operational efficiency in assessing financial competitiveness. This paper adopts a similar analytical framework to evaluate the Indian real estate sector.

According to the World Economic Forum (2008), enterprise competitiveness is defined as "a company's ability to generate more wealth in global markets than its competitors" \cite{wef1984}. In 1990, Prahalad and Hamel, in their seminal work "The Core Competence of the Corporation", emphasized that core competencies represent the collective learning within an organization, particularly in coordinating diverse production skills and integrating multiple technologies. They argued that core competence involves harmonizing streams of technology and focuses on how work is organized, and value is delivered \cite{hbs1990}. Capability theory further suggests that enterprise competitiveness is a capability system where the ability to accumulate, maintain, and develop products and markets is the key to sustaining long-term competitive advantage. Differences in firms' capabilities are critical in determining their competitiveness.

Financial competitiveness research is still in its early stages of exploration. The concept of financial competitiveness is largely derived from the capability theory of enterprise competitiveness. According to Wang Yanhui and Guo Xiaoming\cite{guo}, financial competitiveness refers to the demonstrated ability of a company’s financial operations to achieve its business objectives. It represents a comprehensive strength that arises from integrating financial strategy, financial resources, financial capacity, financial performance, and financial innovation. 

Zhang Youtang and Fen Ziqin \cite{zhang} conceptualize financial competitiveness across three dimensions: the adaptability of financial strategy to the environment, the competitiveness of financial resource allocation, and the integration of financial interests. They propose that financial competitiveness represents the value creation process aligned with the enterprise's strategic direction.

Several scholars have contributed to the concept of financial competitiveness, expanding on its dimensions and strategic importance. Porter (1985) \cite{porter} highlights the significance of financial resource allocation as part of a firm’s overall strategy to achieve competitive advantage. He underscores that effective financial management is critical to value creation and sustaining superior performance in a competitive market. Barney (1991) \cite{doi:10.1177/014920639101700108} emphasizes that financial resources and other key assets are fundamental in achieving long-term competitiveness. The author argues that firms strategically managing their financial resources can better adapt to external changes, positioning themselves for sustained advantage. Grant (1991) further develops this notion, noting that financial resources are essential to a firm’s resource portfolio. He explains that firms that optimise their financial capabilities in line with their strategic goals are better equipped to compete effectively in dynamic markets.

Several studies have explored the evaluation of financial competitiveness, primarily by examining firms' financial indices. The existing literature uses financial indicators to assess various dimensions of financial competitiveness. For instance, Shen Airong \cite{shen} developed a comprehensive evaluation system for financial competitiveness using factor analysis, which incorporated key dimensions such as profitability, debt-paying ability, growth potential, and operational efficiency, measured by 13 selected financial indicators.

Similarly, Altman (1968), in his seminal work on financial ratios, discriminant analysis, and the prediction of corporate bankruptcy, introduced the Z-score model \cite{altman}, which uses financial ratios to assess firms' financial health and competitiveness. His work emphasizes the predictive power of financial metrics like profitability and solvency in determining a firm’s competitive standing.

Additionally, Lev and Sunder (1979) \cite{LEV1979187} highlighted the importance of selecting appropriate financial ratios to measure different dimensions of a firm’s performance. They argued that a well-rounded evaluation system should incorporate multiple financial indicators to capture the various facets of financial competitiveness, similar to Shen’s approach of including profitability, debt-paying capability, and growth capacity.

Wheelen and Hunger (1992) also contributed to the discourse by emphasizing in Strategic Management and Business Policy that firms’ financial performance indicators, such as liquidity, leverage, and profitability, are central to evaluating long-term competitiveness \cite{wheelen1992strategic}. Their framework echoes Shen’s approach, reinforcing the importance of a comprehensive financial metrics evaluation for a company’s competitive strength.

These studies collectively underline the necessity of a multidimensional approach to evaluating financial competitiveness, where various financial indicators provide insights into different aspects of a firm's performance and strategic capacity.

\section{Comprehensive Financial Competitiveness Evaluation Index System}

\subsection{Indicator Selection}

Based on the existing literature and the availability of data, financial competitiveness is typically decomposed into three primary dimensions: profitability, solvency, and operational capacity. These components are commonly used to evaluate a firm's financial performance and competitive standing. In the context of Indian real estate companies, financial competitiveness is assessed using specific metrics categorized into four distinct groups, as outlined in Table \ref{tab:tab1}. Given the capital-intensive nature of the real estate industry, additional capital intensity indicators have been incorporated into the operational metrics section to provide a more comprehensive evaluation.

All the indicator data presented in Table \ref{tab:tab1} were derived and calculated using publicly available information from the annual and quarterly filings of these companies, as reported on the \href{https://www.nseindia.com/}{National Stock Exchange (NSE)} respective to the ticker name. The use of standardized financial data ensures the accuracy and reliability of the evaluation process, providing a solid foundation for further analysis and comparison across companies in the Indian real estate sector.

\begin{table}[t]
\centering
\begin{tabular}{ll}
\toprule
\textbf{Category}                                & \textbf{Indicator}                                              \\ \midrule
\multirow{11}{*}{\textbf{Profitability Metrics}}          & Operating Profit Margin (\%)                                    \\  
                                                 & Profit After Tax for Last 12 Months (in Crores)                 \\ 
                                                 & Return on Capital Employed (\%)                                 \\ 
                                                 & Earnings Before Interest and Tax for Last 12 Months (in Crores) \\ 
                                                 & Net Profit for Last 12 Months (in Crores)                       \\ 
                                                 & Price to Earnings Ratio                                         \\ 
                                                 & Return on Assets for Last 12 Months (\%)                        \\ 
                                                 & Return on Equity (\%)                                           \\  
                                                 & Earnings Yield (\%)                                             \\  
                                                 & Return on Invested Capital (\%)                                 \\ 
                                                 & Break-even Point Percentage                                     \\ \midrule
\multirow{5}{*}{\textbf{Solvency Metrics}}                & Debt (in Crores)                                                \\
                                                 & Debt to Equity Ratio                                            \\ 
                                                 & Interest Coverage Ratio                                         \\ 
                                                 & Debt to Profit Ratio                                            \\ 
                                                 & Leverage Ratio                                                  \\ \midrule
\multirow{7}{*}{\textbf{Sustainable Development Metrics}} & Free Cash Flow (in Crores)                                      \\ 
                                                 & Net Cash Flow (in Crores)                                       \\ 
                                                 & Cash Flow from Operations (in Crores)                           \\ 
                                                 & Dividend Payout Percentage                                      \\ 
                                                 & Capital Employed (in Crores)                                    \\ 
                                                 & Price to Earnings Growth Ratio                                  \\ 
                                                 & Market Price to Free Cash Flow Ratio                            \\ \midrule
\multirow{21}{*}{\textbf{Operational Metrics}}            & Sales Revenue (in Crores)                                       \\  
                                                 & Operating Profit for Last 12 Months (in Crores)                 \\ 
                                                 & Equity Capital (in Crores)                                      \\
                                                 & Capital Work in Progress (in Crores)                            \\ 
                                                 & Current Assets (in Crores)                                      \\
                                                 & Current Liabilities (in Crores)                                 \\ 
                                                 & Total Assets (in Crores)                                        \\ 
                                                 & Working Capital (in Crores)                                     \\ 
                                                 & Inventory (in Crores)                                           \\ 
                                                 & Inventory Turnover Ratio                                        \\ 
                                                 & Quick Ratio                                                     \\ 
                                                 & Asset Turnover Ratio                                            \\
                                                 & Working Capital Days                                            \\ 
                                                 & Cash Conversion Cycle (Days)                                    \\
                                                 & Number of Inventory Days                                        \\ 
                                                 & Days of Receivables                                             \\ 
                                                 & Market Price to Sales Ratio                                     \\
                                                 & Current Ratio                                                   \\ 
                                                 & Working Capital to Sales Percentage                             \\ 
                                                 & Leverage (in Rupees)                                            \\ 
                                                 & Intrinsic Value (in Rupees)                                     \\ \bottomrule
\end{tabular}%

\caption{The Financial Competitiveness Indicators}
\label{tab:tab1}
\end{table}

\subsection{Methodology: The Entropy Method}

Entropy, a fundamental concept in thermodynamics, was first introduced by Rudolf Clausius in 1850 \cite{clausius1850} as a measure of the irreversible dissipation of energy within a system. Clausius’s work laid the foundation for the second law of thermodynamics, which states that the entropy of an isolated system always increases over time. Later, in 1877, Ludwig Boltzmann advanced this concept by linking entropy $S$ with the thermodynamic probability $\Omega$, which represents the number of microscopic quantum states available to a system. Boltzmann formulated this relationship through the equation:

\begin{equation}
    S = K \ ln(\Omega)
\end{equation}

where $K$ is the Boltzmann constant, and $ln(\Omega)$ refers to the natural logarithm of the number of microstates. This equation captures the statistical nature of entropy, associating it with the disorder or randomness at the microscopic level. The more quantum states available, the higher the entropy, signifying a greater level of disorder in the system.

In 1948, Claude E. Shannon \cite{shannon1948communication} introduced the concept of entropy in information theory, defining it as a measure of "information, choice, and uncertainty." Shannon demonstrated that entropy, denoted by $H$, can be mathematically expressed as \cite{shen2010entropy}:

\begin{equation} H = -k \ \sum_{i=1}^{n} p_i \ \log(p_i) \end{equation}

Where k is a positive constant and {$p_1$, $p_2$, . . . , $p_n$} represent the probabilities of a set of possible events. In this context, entropy 
$H$ measures the uncertainty or unpredictability of a system. Specifically, when the distribution of probabilities is more uniform (i.e., the uncertainty is higher), the entropy value increases. Conversely, as the distribution becomes more concentrated (i.e., uncertainty decreases), the entropy value diminishes, reflecting a lower level of uncertainty and a higher amount of information.

In the context of multi-criteria decision-making and financial analysis, entropy is used to evaluate the degree of variability or dispersion among indicators. As per Shannon's entropy theory, the higher the entropy associated with a particular indicator, the greater its variability, and consequently, its weight in the overall evaluation increases. This principle is particularly useful for financial competitiveness evaluation, where multiple financial metrics, such as profitability, liquidity, and solvency, are considered. The greater the variability of these indicators, the more significant their contribution to the comprehensive evaluation.

According to the basic principle of the entropy method, the evaluation of financial competitiveness is as follows: 
\begin{itemize}
    \item \textit{\textbf{Non-dimensionalization of indicators}} \\ \\

    The evaluation of financial competitiveness indicators across companies often involves metrics that are expressed in different units of measurement, which can introduce inconsistencies in comparative analyses. To mitigate this issue, it is essential to normalize or non-dimensionalize the data, eliminating the negative effects that arise from unit disparities. Non-dimensionalization not only facilitates a consistent evaluation across different indicators but also accounts for the directionality of each indicator. 

For instance, when an indicator is considered positive — meaning that higher values indicate better performance, such as in profitability or return on assets—the non-dimensionalized value of indicator $j$ for company $i$, denoted as $s_{ij}$, is calculated using the following formula:

\begin{equation}
    s_{ij} = \frac{r_{ij} - \min(r_{j})}{\max(r_{j}) - \min(r_{j})}
\end{equation}

In this equation, $r_{ij}$ represents the original value of the $j^{th}$ indicator for the $i^{th}$ company, while $\max(r_{j})$ and $\min(r_{j})$ denote the maximum and minimum values of the indicator across all companies, respectively. This formula ensures that all positive indicators are scaled between 0 and 1, with 1 representing the best performance.

Conversely, for inverse indicators, where lower values reflect better performance, such as in debt ratios or operational costs, the non-dimensionalization process is computed differently to reflect this inverse relationship. The non-dimensionalized value $s_{ij}$ for an inverse indicator is calculated as:

\begin{equation}
    s_{ij} = \frac{\min(r_{j}) - r_{ij}}{\max(r_{j}) - \min(r_{j})}
\end{equation}

This adjustment ensures that for inverse indicators, lower raw values are transformed into higher non-dimensionalized scores, thus aligning with the overall objective of evaluating financial competitiveness.

Non-dimensionalization is a critical step in ensuring that the evaluation process is both fair and accurate, especially when dealing with composite indicators across a diverse set of firms or industries. As noted by \cite{e24081056} and \cite{guo}, proper normalization enables comparability and strengthens the validity of composite evaluations, particularly in capital-intensive industries such as real estate or manufacturing, where diverse financial metrics are involved.


    \item \textit{ \textbf{Estimation of Cumulative Distribution Function}} \\ 

    Let $R_j$ represent the set of data points for indicator $j$ across all companies. To estimate the distribution of indicator $j$, we employ Kernel Density Estimation (KDE), a non-parametric technique that helps in deriving the probability density function (PDF) of a given dataset without making any assumptions about its underlying distribution. This method is especially useful in financial research where data distributions are often skewed or exhibit heavy tails. \\

Once the PDF is estimated, the cumulative distribution function (CDF), denoted as $\varphi_j(x)$, is derived. The CDF represents the probability that a randomly chosen data point from the distribution will be less than or equal to a certain value $x$. One of the key properties of the CDF is that it is a monotonically increasing function, meaning that as the value of $x$ increases, so does the cumulative probability, ensuring that the function progresses in only one direction. This is expressed mathematically as:

\begin{equation}
    0 \ \leq \ \varphi_j(x) \ \leq \ 1
\end{equation}

This property implies that the CDF will always lie between 0 and 1, where $\varphi_j(x) = 0$ when $x$ takes its minimum value and $\varphi_j(x) = 1$ when $x$ reaches its maximum value within the dataset. This boundedness allows for a straightforward interpretation of the results, which is essential when comparing multiple financial indicators across companies.

    \item \textit{\textbf{Computation of Entropy}} \\ 

    Entropy, as traditionally defined, is calculated based on the probabilities of discrete events, as shown in Shannon's  entropy formula \cite{shannon1948communication}. However, in the context of financial indicators used in this research, the data are not discrete but continuous, requiring an adaptation of the traditional entropy framework. To accommodate this, we propose a continuous form of entropy tailored to the characteristics of our financial indicators, which aligns with the methodology for continuous data evaluation. \\

The continuous entropy for a given indicator $j$ is defined as follows:

\begin{equation}
    H_j = - \int_{0}^{1} \varphi_j (x) \ ln(\varphi_j (x)) \ dx
\end{equation}

where $\varphi_j(x)$ represents the cumulative distribution function (CDF) of indicator $j$, and the constant $e$ is the base of the natural logarithm. The integral is evaluated over the interval $[0, 1]$, reflecting the normalized nature of the CDF. \\

The use of the logarithmic term ensures that values of $\varphi_j(x)$ closer to 1 (indicating a higher concentration of data in the distribution) contribute less to the overall entropy value, signifying lower uncertainty. Conversely, values of $\varphi_j(x)$ closer to 0 indicate greater dispersion in the data, contributing more to the entropy and, in turn, suggesting higher uncertainty. This characteristic allows the entropy measure to effectively capture the degree of variability or uncertainty within the distribution of each financial indicator.

    \item \textit{\textbf{Computation of the weight for each indicator}} \\  \\
    The weight assigned to each financial indicator $j$, denoted as $w_j$, is calculated using the entropy values derived from the continuous entropy model. The weight reflects the relative importance of each indicator based on its entropy value, which measures the degree of uncertainty or variability associated with the indicator's distribution.\\

The weight for indicator $j$ is given by the formula:

\begin{equation}
    w_j = \frac{H_j}{\sum_{k=1}^{m} H_k}
\end{equation}

where $H_j$ is the entropy value of the $j^{th}$ indicator, and $m$ represents the total number of indicators included in the analysis. The denominator, $\sum_{k=1}^{m} H_k$, is the sum of entropy values across all $m$ indicators, ensuring that the weights sum to 1.\\

This approach ensures that indicators with higher entropy values, which exhibit greater dispersion or variability, are given more weight in the overall evaluation process. In contrast, indicators with lower entropy values, indicating less variability and thus less influence on differentiating the financial competitiveness of firms, are assigned lower weights. The method provides an objective, data-driven way to assign weights, avoiding the subjective biases that can arise from manually assigning importance to indicators.

    \item \textit{\textbf{Evaluation of Financial Competitiveness }} \\ 

    The integrated score provides a comprehensive evaluation of the financial competitiveness of each company. This score, denoted as $F_i$, is computed as a weighted sum of the non-dimensionalized financial indicators $s_{ij}$ for company $i$, with each indicator $j$ assigned a weight $w_j$ based on its entropy-derived importance. The formula for the integrated financial competitiveness score is:

\begin{equation}
     F_i = \sum_{j=1}^{n} w_j s_j \  \ \ \ \ \ \ for \ \  i = 1, 2, ..., n
\end{equation}

where $n$ represents the total number of indicators, and $w_j$ is the weight associated with each indicator $j$, as determined through entropy-based weighting. \\

This integrated score reflects the overall financial performance of each company across multiple dimensions, including profitability, solvency, sustainable development and operational capacity. By utilizing non-dimensionalized data, this approach ensures comparability across indicators with different units of measurement, thus eliminating biases that could arise from inconsistencies in the scale of financial metrics.

\end{itemize}

\section{Empirical Results and Analysis}

\subsection{Sample Selection \& Data Consolidation}
This study incorporates financial data from over 50 publicly listed Indian real estate companies, encompassing a wide spectrum of firms from large multinational conglomerates to smaller regional developers. By including companies of varying sizes and market reach, the study offers a comprehensive analysis of the financial health and competitiveness of the Indian real estate industry. This broad representation allows for a more nuanced understanding of the sector, highlighting both industry-wide trends and the financial dynamics unique to different types of firms.

The financial data used in this study is sourced from publicly available filings, including annual reports, quarterly earnings statements, and other regulatory disclosures submitted to the \href{https://www.nseindia.com/}{National Stock Exchange (NSE)}. These filings provide detailed information on key financial indicators such as profitability, solvency, operational capacity, sustainable development and cash flow, which are essential for evaluating the financial competitiveness of each company.

\subsection{Determining the Weights }
By employing the entropy method, the weights for each financial indicator were calculated based on the data collected from the selected Indian real estate companies. In this context, the entropy method measures the degree of dispersion across data points, assigning higher weights to indicators with greater variability, thereby reflecting their significance in differentiating company performance.

Notably, metrics like EBITDA, PE Ratio, and Enterprise Value received the highest weights, underscoring their critical role in determining financial competitiveness within the real estate sector. These indicators reflect a company’s profitability and market valuation, crucial factors for evaluating performance in capital-intensive industries like real estate. This emphasis on key financial metrics is consistent with findings from prior research, which highlights their importance in assessing both short-term profitability and long-term growth potential.

\begin{table}[h]
\centering
\begin{tabular}{llll|lll}
\toprule
\textbf{Rank} & \textbf{NSE Ticker Name}        & \textbf{Score} & \textbf{} & \textbf{Rank} & \textbf{NSE Ticker Name} & \textbf{Score} \\
\midrule
\textit{1}  & DLF        & 0.09106 &  & \textit{26} & PURVA       & 0.07704 \\
\textit{2}  & OBEROIRLTY & 0.08843 &  & \textit{27} & ARVSMART    & 0.07701 \\
\textit{3}  & LODHA      & 0.08641 &  & \textit{28} & PENINLAND   & 0.07700 \\
\textit{4}  & PRESTIGE   & 0.08367 &  & \textit{29} & ASAL        & 0.07690 \\
\textit{5}  & DBREALTY   & 0.08338 &  & \textit{30} & HAZOOR      & 0.07685 \\
\textit{6}  & PHOENIXLTD & 0.08294 &  & \textit{31} & MAXESTATES  & 0.07674 \\
\textit{7}  & EMBASSY    & 0.08258 &  & \textit{32} & ASHIANA     & 0.07670 \\
\textit{8}  & GODREJPROP & 0.08152 &  & \textit{33} & KEYSTONE    & 0.07669 \\
\textit{9}  & NEXUSSEL   & 0.08105 &  & \textit{34} & SHRIRAMPPS  & 0.07663 \\
\textit{10} & MINDSPACE  & 0.08067 &  & \textit{35} & SOBHA       & 0.07660 \\
\textit{11} & NESCO      & 0.07949 &  & \textit{36} & NATIONALUM  & 0.07657 \\
\textit{12} & NIRLON     & 0.07914 &  & \textit{37} & SUMIT       & 0.07655 \\
\textit{13} & BRIGADE    & 0.07903 &  & \textit{38} & NILAINFRA   & 0.07643 \\
\textit{14} & BIRET      & 0.07823 &  & \textit{39} & SIGNATUREGL & 0.07627 \\
\textit{15} & ANANTRAJ   & 0.07822 &  & \textit{40} & TEXINFRA    & 0.07586 \\
\textit{16} & MARATHON   & 0.07781 &  & \textit{41} & TARC        & 0.07580 \\
\textit{17} & ELPROINTL  & 0.07774 &  & \textit{42} & KOLTEPATIL  & 0.07552 \\
\textit{18} & SBGL       & 0.07767 &  & \textit{43} & MAHLIFE     & 0.07532 \\
\textit{19} & GEECEE     & 0.07747 &  & \textit{44} & HUBTOWN     & 0.07486 \\
\textit{20} & NEWINFRA & 0.07745        &           & \textit{45}   & OMAXE                    & 0.07338        \\
\textit{21} & AJMERA     & 0.07745 &  & \textit{46} & PARSVNATH   & 0.07325 \\
\textit{22} & ELDEHSG    & 0.07720 &  & \textit{47} & PVP         & 0.07313 \\
\textit{23} & SUNTECK    & 0.07713 &  & \textit{48} & EMBDL       & 0.07257 \\
\textit{24} & RDBRIL     & 0.07712 &  & \textit{49} & UNITECH     & 0.04935 \\
\textit{25} & AGI        & 0.07709 &  & \textit{50} & HEMIPROP    & 0.01893 \\
\bottomrule
\end{tabular}
\caption{Result of the financial competitiveness calculation}
\label{tab:tab2}
\end{table}

\subsection{Result Evaluation}

\begin{figure}
    \centering
    \includegraphics[width=\linewidth]{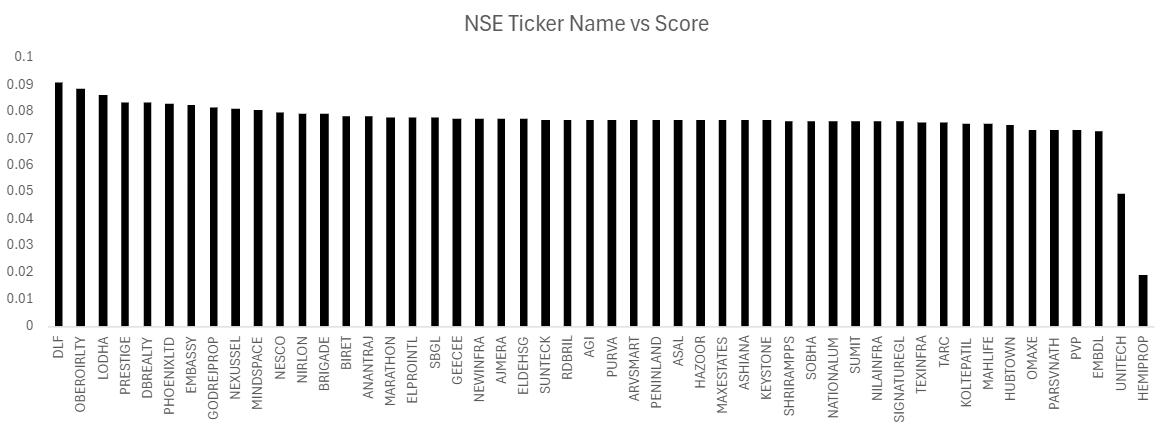}
    \caption{Competitive Score of companies}
    \label{fig:chart}
\end{figure}

The final financial competitiveness scores for Indian real estate companies are displayed in Table \ref{tab:tab2}
, which includes the corresponding ranks and scores as illustrated in Figure \ref{fig:chart}. These scores provide a comprehensive evaluation of each firm's financial strength and competitive positioning within the sector. The ranking methodology is based on a multi-dimensional analysis, incorporating key financial indicators such as profitability, debt management, operational efficiency, and growth potential.

The top-ranked company achieved a normalized score 1 for key financial metrics, including Profit After Tax (PAT), Net Profit, Total Assets, and Cash Flow from Operations, reflecting its strong financial performance. Additionally, its Intrinsic Value (0.964), Leverage Ratio (0.993), and Operating Profit Margin (0.980) were also near 1, indicating robust financial health and operational efficiency.

The second-ranked company excelled in its Capital Work in Progress, which reached a normalized score of 1. Its Leverage Ratio and Operating Profit Margin were also close to 1, demonstrating strong capital management and operational profitability.

In contrast, the lowest-ranked company had a Cash Conversion Cycle and debt-profit ratio normalized at 1, indicating that these metrics contributed minimally to its overall financial competitiveness. This suggests inefficiencies in managing working capital and high debt levels relative to profit, adversely affecting its competitive positioning.

\subsection{Correlation analysis}

This research aims to develop a financial analysis method that effectively captures the core competitiveness of enterprises. So these indices aim to clearly represent an enterprise’s competitive standing. Given that the primary characteristics of core competitiveness are taken from 33 indicators to some extent, interrelated, the correlation between various variables is acknowledged and deemed acceptable within the scope of this analysis as shown in Table \ref{tab:corr} for the first four indicators in Profitability Metrics as shown in Table \ref{tab:tab1} in which $X_1$ is Operating profit Margin and so on for $X_2$, $X_3$ and $X_4$. \\

\begin{table}
\centering
\begin{tabular}{l|l}
\toprule
Mean     & 0.039346  \\ 
Median   & 0.036645  \\
Std. Dev & 0.014627  \\
Kurtosis & 20.29681  \\
Skewness & 3.902656  \\
Smallest & 0.012932       \\
Largest  & 0.120858        \\
Obs      & 52 \\
\bottomrule
\end{tabular}
\caption{Descriptive Statistics Of Score}
\label{tab:my-table}
\end{table}

The correlation analysis revealed that the Debt to Profit Ratio and the Cash Conversion Cycle exhibit a similar distribution pattern, indicating a significant relationship between these variables. Furthermore, both Profit After Tax for the Last 12 Months and Net Profit for the Last 12 Months were highly correlated, with correlation coefficients approaching 0.99 in both cases. This near-perfect correlation suggests that these metrics are almost identical in reflecting the financial performance of the enterprises under consideration, reinforcing their importance in evaluating overall profitability and financial health.

Such findings are consistent with prior research, such as Chen and Zhang (2018) \cite{zhang}, who found that profitability metrics like net profit and post-tax earnings often exhibit high correlations due to their shared reliance on underlying revenue and cost structures. 

\begin{table}
\centering
\begin{tabular}{l|llll}
\toprule
     & $X_1$       & $X_2$       & $X_3$       & $X_4$ \\
     \midrule
$X_1$ & 1           &             &             &       \\
$X_2$ & 0.160421117 & 1           &             &       \\
$X_3$ & 0.214680333 & 0.360263135 & 1           &       \\
$X_4$ & 0.130978729 & 0.615259414 & 0.041744418 & 1 \\
\bottomrule
\end{tabular}
\caption{Indicator Correlation Coefficients}
\label{tab:corr}
\end{table}

\section{Conclusion}

Research on financial competitiveness remains exploratory, with limited literature available. The financial competitiveness of companies varies significantly across industries, making it challenging to develop a universal evaluation system applicable to all firms. This study developed a comprehensive financial competitiveness evaluation system for real estate companies. Based on the conceptual framework of financial competitiveness and the unique characteristics of the real estate sector, the evaluation system is structured into four key dimensions: profitability, solvency, sustainable development, and operational capacity.

Using the principle of information entropy, an objective method was proposed to determine the weight of each indicator within the system. This analysis reveals that high-scoring companies exhibit strong profitability and robust operational capacity and maintain high leverage, which collectively enhance their financial competitiveness. In contrast, low-scoring companies demonstrate weaker profitability and poor sustainable development capacity. Interestingly, metrics such as the debt-to-profit ratio and the cash conversion cycle were found to have minimal impact on the overall financial competitiveness of these firms.

These findings align with earlier research by Porter (1985) \cite{porter} and Barney (1991) \cite{doi:10.1177/014920639101700108}, which emphasize the critical role of profitability, leverage, and operational efficiency in maintaining a competitive edge. However, the limited influence of the debt-to-profit ratio and cash conversion cycle highlights the unique financial dynamics of the real estate industry, where long-term asset growth and capital-intensive operations may play a more significant role. This study contributes to the broader discourse on financial competitiveness by offering a tailored evaluation model for the real estate sector, which can serve as a foundation for future research and industry-specific analysis.

\bibliographystyle{unsrt}
\bibliography{references}







\end{document}